\documentclass{aa}
\usepackage{graphicx}
\usepackage{txfonts}
\begin{document}
\title{A Spectroscopic study of the non-thermal radio emitter Cyg\,OB2\,\#8A: Discovery of a new binary system\thanks{Based on observations collected at the Observatoire de Haute Provence, France.}}
\author{M.\ De Becker \and G.\ Rauw\thanks{Research Associate FNRS (Belgium)} \and J. Manfroid\thanks{Research Director FNRS (Belgium)}} 
\offprints{M. De Becker}
\institute{Institut d'Astrophysique et de G\'eophysique, Universit\'e de Li\`ege,
17, All\'ee du 6 Ao\^ut, B5c, B-4000 Sart Tilman, Belgium}

\date{Received date / Accepted date}
\authorrunning{M.\ De Becker et al.}
\titlerunning{Cyg\,OB2\,\#8A: Discovery of a new binary system}

\abstract{We present the results of a spectroscopic campaign revealing that the non-thermal radio emitter Cyg\,OB2\,\#8A is an O6 + O5.5 binary system. We propose the very first orbital solution indicating a period of about 21.9 days. The system appears to be eccentric (0.24 $\pm$ 0.04) and is likely seen under a rather low inclination angle. The mass ratio of the components is close to unity. The impact of the binarity of this star in the framework of our understanding of non-thermal radio emission from early-type stars is briefly discussed.  
\keywords{binaries: spectroscopic -- stars: early-type -- stars: fundamental parameters  -- stars: individual: Cyg\,OB2\,\#8A}}
\maketitle

\section{Introduction}
Cyg\,OB2 is one of the most interesting OB association of our Galaxy. It is known to harbour a huge number of OB stars, among which about 100 O-stars (Kn\"odlseder \cite{kn}; Comer\'on et al.\,\cite{Com}). Considering its mass, density and size, Kn\"odlseder (\cite{kn}) suggested to re-classify it as a young globular cluster, the first object of this class in our Galaxy. However, the extremely large and non-uniform reddening towards Cyg\,OB2 renders its study at optical wavelengths extremely difficult.\\
One of the optically brightest O-stars in Cyg\,OB2 is Cyg\,OB2\,\#8A (BD\,+40$^\circ$\,4227, m$_\mathrm{V}$=9.06). This star is classified as O5.5I(f) (Massey \& Thompson\,\cite{MT}). Cyg\,OB2\,\#8A is one of the first early-type stars discovered to be an X-ray emitter with {\it Einstein} (Harnden et al.\,\cite{har}). Its X-ray emission was subsequently investigated with {\it ROSAT} (Waldron et al.\,\cite{wal}) and {\it ASCA} (Kitamoto \& Mukai \cite{KM}; De Becker \cite{DeBMT}).\\
A peculiarity of this star is that it is a non-thermal radio emitter (Bieging et al.\,\cite{BAC}). This non-thermal emission requires the existence of a population of relativistic electrons, believed to be accelerated in shocks through the Fermi mechanism (Chen \& White \cite{CW}). An important question is what is the nature of the shocks responsible for this acceleration process: intrinsic shocks due to radiative instabilities within the wind of a single star (e.g. Chen \& White\,\cite{CW}), or a wind-wind collision within binary systems (e.g. Eichler \& Usov \,\cite{EU})? Concerning the latter scenario we note that while the majority of non-thermal radio emitting Wolf-Rayet stars are indeed binaries (Dougherty \& Williams \cite{DW}), the situation is less clear for O-stars (De Becker et al.\,\cite{DeB2}). In particular, up to now no evidence for binarity was put forward for Cyg\,OB2\,\#8A. For instance, Lortet et al. (\cite{Lor}) did not detect any bright ($\Delta$m $\leq$ 3\,mag) visual companion with an angular separation in the range 0.04 to 1.5 arcsec.\\
In this paper, we present the first spectroscopic monitoring of Cyg\,OB2\,\#8A revealing that it is a binary system. The observations are described in Sect.\,\ref{obs}. The spectrum is discussed in Sect.\,\ref{specan}, and in Sect.\,\ref{binary} we present our period determination, as well as the first orbital solution proposed for this system. The conclusions and prospects are finally given in Sect.\,\ref{concl}.

\section{Observations and data reduction \label{obs}}
Spectroscopic observations of Cyg\,OB2\,\#8A were collected at the Observatoire de Haute-Provence (OHP, France) during four observing runs in September 2000, September 2001, September 2002 and October 2003. All observations were carried out with the Aur\'elie spectrograph fed by the 1.52\,m telescope (Gillet et al.\,\cite{Gil}). Aur\'elie was equipped with a 2048$\times$1024 CCD EEV 42-20\#3, with a pixel size of 13.5 $\mu$m squared. All spectra were taken with a 600 l/mm grating with a reciprocal dispersion of 16 \AA\,mm$^{-1}$, allowing us to achieve a spectral resolution of about 8000 in the blue range.

A total of 35 spectra were obtained over the wavelength range between about 4455 and 4900 \AA\,. The data were reduced using the {\sc midas} software developed at ESO, with the same procedure as described by Rauw \& De Becker (\cite{ic1805}).  Table\,\ref{log} gives the journal of our observations, including the radial velocity measured on the \ion{He}{i} $\lambda$ 4471 line (see Sect.\,\ref{per}). The mean signal-to-noise ratio of individual spectra of our data set, estimated over a region devoid of lines, is about 250.

\begin{table}
\caption{Journal of our observations of Cyg\,OB2\,\#8A. The first column provides the heliocentric julian date (in the format HJD -- 2\,450\,000). The second column gives the phase according to our orbital solution (see Table\,\ref{par}). The third and fourth columns yield the radial velocities measured on the \ion{He}{i} $\lambda$ 4471 line for the two components of the binary system. The last column gives the weight assigned to the various RV values in the orbital solution. The RVs with a weight superior to 0.5 are those which were determined through the simultaneous fit of two Gaussians with the {\it simplex method}.\label{log}}
\begin{center}
\begin{tabular}{l c r r c}
\hline\hline
Date & Phase & RV$_\mathrm{1}$ & RV$_\mathrm{2}$ & Weight \\
 &  & (km\,s$^{-1}$) & (km\,s$^{-1}$) &  \\
\hline
1810.466 & 0.152 &	 23.4	&	--74.6 & 0.1 \\
1811.468 & 0.198 &	 55.2	&	--97.0 & 0.1 \\
1812.475 & 0.244 &	 36.3	&	--110.6 & 0.3 \\
1813.523 & 0.291 &	54.7	&	--84.7 & 0.3 \\		
1814.510 & 0.336 &	 59.9	&	--111.1 & 0.3 \\
1815.503 & 0.382 &	 56.3	&	--121.5 & 1.0 \\
1819.462 & 0.563 &	 34.7	&	--86.1 & 0.3 \\ 
1821.488 & 0.655 &	10.1	&	--58.9 & 0.1 \\
2163.394 & 0.261 &	 39.4	&	--97.2 & 0.5 \\
2164.382 & 0.307 &	 53.1	&	--104.4 & 1.0 \\
2165.361 & 0.351 &	 72.3	&	--80.5 & 1.0 \\
2167.367 & 0.443 &	 53.7	&	--84.4 & 0.7 \\
2168.376 & 0.489 &	 38.7	&	--44.5 & 0.3 \\
2169.371 & 0.535 &	 17.0	&	--32.6 & 0.1 \\ 
2170.351 & 0.579 &	  0.7	&	--42.9 & 0.1 \\
2170.446 & 0.584 &	  0.5	&	--33.0 & 0.1 \\
2518.409 & 0.466 &	 29.9	&	--90.7 & 0.1 \\
2520.369 & 0.556 &	 21.4	&	--76.4 & 0.1 \\
2522.361 & 0.647 &	 --1.2 &	--44.8 & 0.1 \\
2524.354 & 0.738 &	 --60.8	&	34.2 & 0.1 \\
2529.394 & 0.968 &	 --110.5 &	65.2 & 1.0 \\
2532.340 & 0.103 &	  6.9 &		--54.8 & 0.1 \\
2532.421 & 0.106 &	  9.4 &		--25.5 & 0.1 \\
2533.318 & 0.148 &	 18.8 &		--87.1 & 0.1 \\
2533.432 & 0.153 &	 21.1 &		--81.5 & 0.1 \\
2919.402 & 0.771 &	 --66.0	&	41.2 & 0.1 \\
2922.388 & 0.907 &	 --85.9 &	114.4 & 1.0 \\
2923.336 & 0.950 &	 --113.6 &	84.2 & 1.0 \\
2925.407 & 0.045 &	 --50.4	& 	17.9 & 0.05 \\
2926.671 & 0.103 &	 --43.9	&	11.1 & 0.1 \\
2928.349 & 0.179 & 	 37.0	&	--84.4 & 0.7 \\
2929.283 & 0.221 &	 59.0	&	--83.1 & 0.3 \\
2930.334 & 0.270 &	 36.6	&	--91.4 & 0.3 \\
2934.286 & 0.450 &	 62.2	&	--102.0 & 0.7 \\
\hline
\end{tabular}
\end{center}
\end{table}

\section{Spectral analysis \label{specan}}
The inspection of our time series of spectra reveals that some lines display variations pointing towards Cyg\,OB2\,\#8A being a binary system. This is most obvious for the \ion{He}{i} $\lambda$ 4471 and \ion{He}{ii} $\lambda$ 4686 lines. The former shows indeed a (partial) deblending of lines compatible with a binary system, and the latter undergoes strong variability even if no clear deblending is observed.

\begin{figure}[ht]
\begin{center}
\resizebox{8.5cm}{4.0cm}{\includegraphics{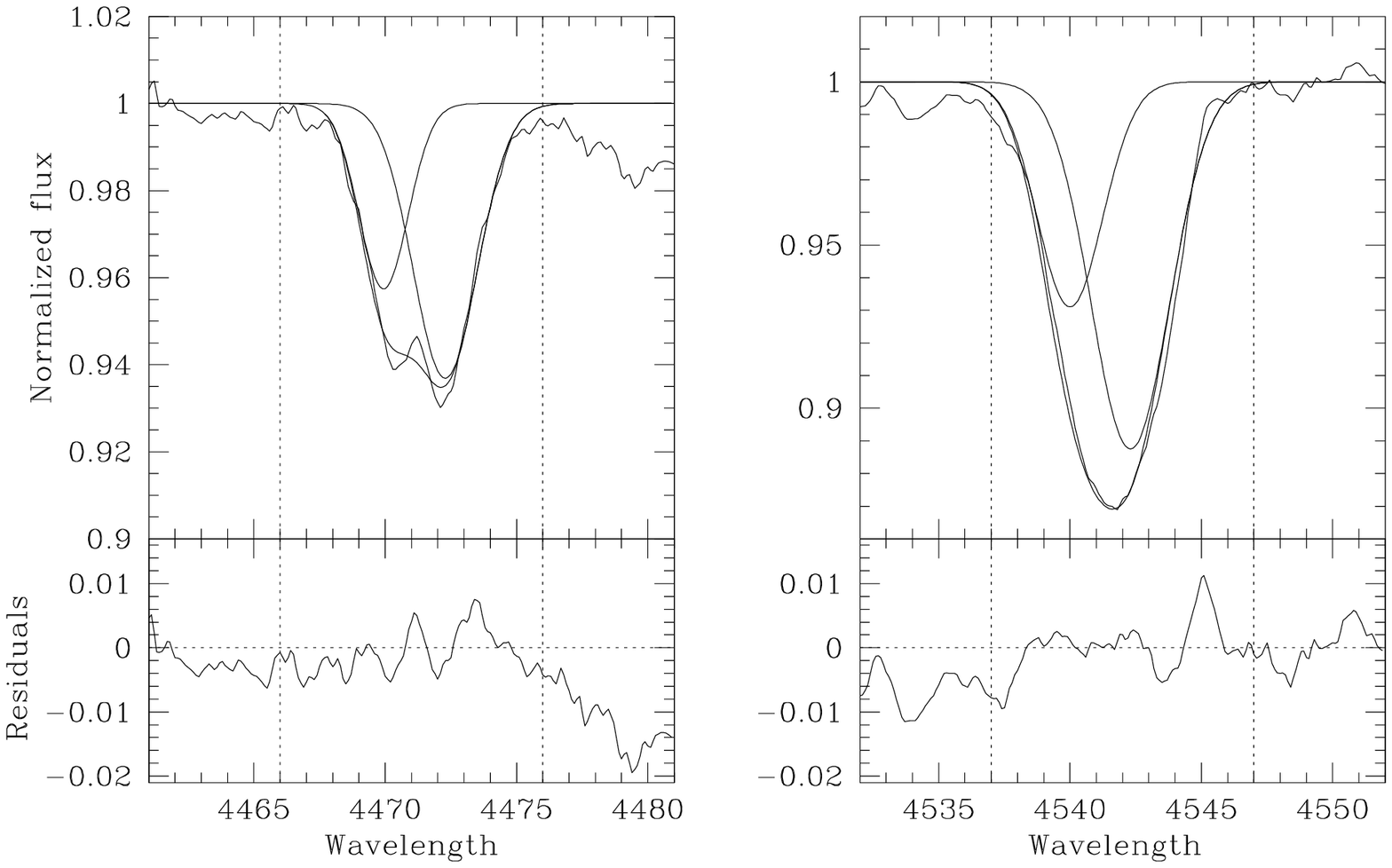}}
\resizebox{8.5cm}{4.0cm}{\includegraphics{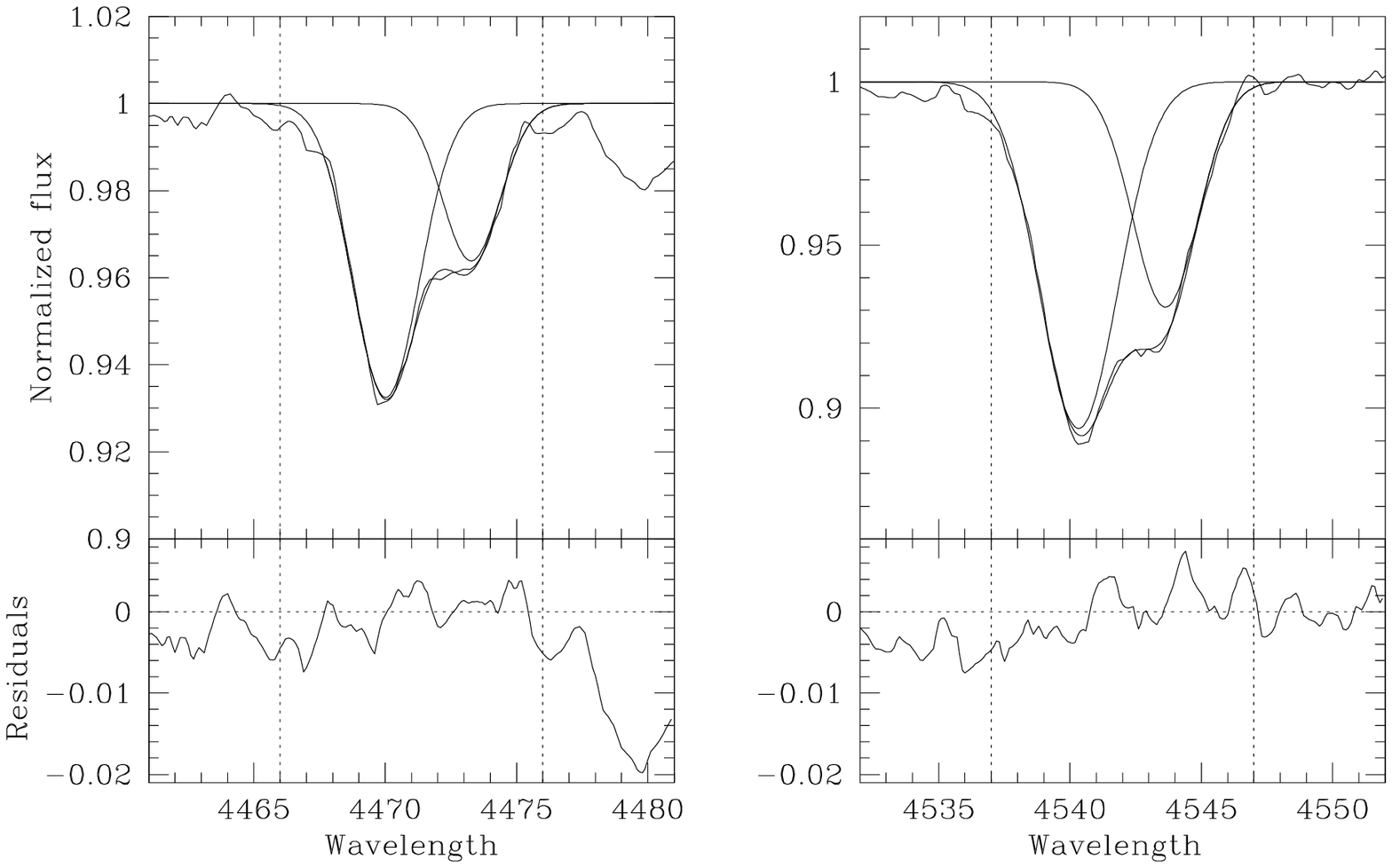}}
\end{center}
\caption{Results of the fit of two Gaussians to the \ion{He}{i} $\lambda$ 4471 (left) and \ion{He}{ii} $\lambda$ 4542 (right) lines for the spectra obtained on HJD 2\,452\,164.382 and HJD 2\,452\,922.388 respectively for the upper and lower figures. These profiles correspond respectively to phases 0.307 and 0.907 ouf our orbital solution. The individual components are overplotted. The vertical dashed lines give the boundaries of the domain where the fit was performed. The lower panels provide the residuals of the fit.\label{fit}}
\end{figure}

We measured the equivalent widths (EWs) of the \ion{He}{i} $\lambda$ 4471 and \ion{He}{ii} $\lambda$ 4542 lines to determine the spectral types of both components. To separate the lines of the two stars, we performed the measurements on the spectra displaying the most pronounced separation between the components (HJD 2\,452\,922.388). We simultaneously fitted two Gaussians to the profile, allowing the central position, the width, and the normalization factor of both lines to vary. The best fit parameters were obtained using an N-dimensional {\it downhill simplex method} (Nelder \& Mead \cite{NM}), with N being the number of free parameters (6 in our case). The integration of the two individual Gaussians gives the EWs of the lines of the components of the binary. Figure \ref{fit} shows the results obtained for the \ion{He}{i} $\lambda$ 4471 and \ion{He}{ii} $\lambda$ 4542 lines at the date near quadrature selected for the spectral type determination. Using the classification criterion of Mathys (\cite{Mat}), the EW ratios obtained from these fits allow us to infer O6 and O5.5 spectral types respectively for the primary and the secondary. We note that spectra obtained at the opposite quadrature display similar intensities for both components. For instance, the profiles displayed in Fig.\,\ref{fit} give intensities (as well as widths and consequently EWs) which do not differ by more than about 5-6\%. Consequently, we do not find evidence for a significant `Struve-Sahade' effect (see e.g. Bagnuolo et al.\,\cite{bag}). The intensities of the lines in the spectra of the two components roughly yield a visual brightness ratio of 2 between the primary and the secondary. Since the spectral types of both components are similar, this suggests that their luminosity classes might be different. With an O6I + O5.5III binary system, the large M$_V$ mentioned by Herrero et al. (\cite{HPN}) is readily explained.

\section{Cyg\,OB2\,\#8A as a binary system \label{binary}}
\subsection{Period determination \label{per}}
As a first step, we searched for the period of the system through a study of the line profile variability of our time series. A Time Variance Spectrum (TVS, Fullerton et al.\,\cite{Ful}) computed between 4455 and 4890 \AA\, on all our spectra reveals a significant variability for \ion{He}{i} $\lambda$ 4471, \ion{He}{ii} $\lambda$$\lambda$ 4542,4686, H$\beta$ and \ion{N}{iii} $\lambda$$\lambda$ 4634,4641. In each case, the TVS displays a double peaked profile suggesting a simultaneous variability of the wings of these lines, compatible with a binary scenario. We performed a Fourier analysis on our times series using the method described by Heck et al. (\cite{HMM}), and used e.g. by Rauw \& De Becker (\cite{ic1805}) for BD\,+60$^\circ$\,497. All periodograms display peaks at frequencies of 0.043 and 0.094 d$^{-1}$, except for the \ion{He}{ii} $\lambda$ 4686 line which does not show the 0.094 d$^{-1}$ peak. Prewhitening of the periodogram (see e.g. Rauw et al.\,\cite{Rauw}) reveals that these frequencies are harmonics, corresponding respectively to $P=23.26$ d and about $P/2$. The former period is more compatible with our data even though $P/2$ gives often the strongest peak. This could be explained by the typical duration of our observing runs (max. 16 d), shorter than the actual period.

As a second step, we measured the radial velocities of both components of the system\footnote{One of our spectra was dropped because of a strong contamination of the \ion{He}{i} $\lambda$ 4471 line profile by a cosmic.} using the fitting method described in the previous section when the separation of the lines was sufficient. When the lines were too heavily blended, we set the line widths and intensities to their values determined from the fits near quadrature and we used an iterative scheme where the line positions were first estimated by eye and subsequently changed to improve the residuals. These measurements were performed only for the \ion{He}{i} $\lambda$ 4471 line, as the other lines were not sufficiently separated. The typical error on the RVs is about 10 km\,s$^{-1}$ for the cases where we were able to fit Gaussians\footnote{This error on the radial velocity corresponds to the standard deviation determined for the radial velocity of a Diffuse Interstellar Band (DIB) at about 4762 \AA\, obtained for spectra of similar S/N ratio.}, and we estimate that it can reach about 25-35 km\,s$^{-1}$ at phases where the lines are severely blended. The Heck et al. (\cite{HMM}) Fourier technique was then applied to our RV time series of the primary (RV$_\mathrm{1}$), the secondary (RV$_\mathrm{2}$), and to the difference RV$_\mathrm{1}$--RV$_\mathrm{2}$. The periodograms of the RV$_\mathrm{1}$ and RV$_\mathrm{2}$ series present both highest peaks corresponding to periods of 23.31 and 21.88 d ($\pm$ 0.04 d). These values emerge also from the RV$_\mathrm{1}$--RV$_\mathrm{2}$ data set. However, we have folded our RVs with these two periods with some success and both values were used as first guesses in our search for an orbital solution.

\subsection{Orbital solution \label{os}}
\begin{figure}
\begin{center}
\resizebox{8.5cm}{4.5cm}{\includegraphics{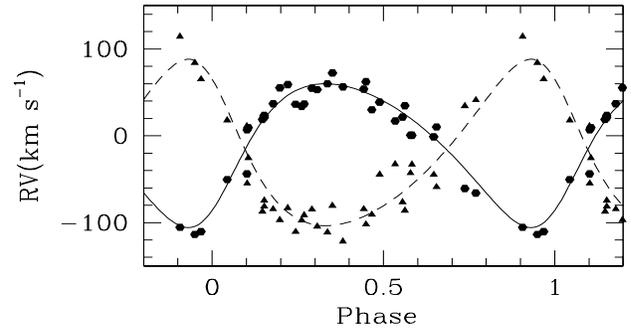}}
\end{center}
\caption{Radial velocity curve of Cyg\,OB2\,\#8A for an orbital period of 21.908 d. The hexagons (resp. triangles) stand for the primary (resp. secondary) RVs. The solid and dashed lines yield our best fit orbital solution respectively for the primary and the secondary.\label{orb}}
\end{figure}

We obtained the first orbital solution for this system using the method described by Sana et al. (\cite{sana}), and used e.g. by Rauw \& De Becker (\cite{ic1805}). We assigned various weights to our data to take into account the different errors affecting our RV measurements. Due to the intensity ratio of the lines of the two components ($\sim$ 2), there was no ambiguity on the identification of the lines respectively of the primary and the secondary. Table\,\ref{par} yields the main parameters of the system according to our best orbital solution\footnote{The orbital solution with the period of about 23.3 d turned out to be incompatible with a spectrum kindly provided by Dr. A. Herrero (private communication).}. As can be seen from Fig.\,\ref{orb}, the system is eccentric with $e$ = 0.24 $\pm$ 0.04. However, we emphasize that the error we obtain on the eccentricity underestimates the actual error as a result of the rather heterogeneous phase coverage of our time series (see Fig.\,\ref{orb}).

\begin{table}
\caption{Orbital solution for Cyg\,OB2\,\#8A. T$_{\circ}$ refers to the time of periastron passage. $\gamma$, $K$, and $a\,\sin\,i$ denote respectively the systemic velocity, the amplitude of the radial velocity curve, and the projected separation between the centre of the star and the centre of mass of the binary system. R$_\mathrm{RL}$ stands for the radius of a sphere with a volume equal to that of the Roche lobe computed according to the formula of Eggleton (\cite{Eg}).\label{par}}
\begin{center}
\begin{tabular}{l c c }
\hline\hline
  & Primary & Secondary \\
\hline
P (days)  & \multicolumn{2}{c}{21.908 (fixed)} \\
$e$   & \multicolumn{2}{c}{0.24 $\pm$ 0.04} \\
T$_\circ$ (HJD--2\,450\,000) & \multicolumn{2}{c}{1807.139 $\pm$ 0.894} \\
$\gamma$ (km\,s$^{-1}$)  & --8.1 $\pm$ 3.3 & --25.0 $\pm$ 3.6 \\
$K$ (km\,s$^{-1}$) & 82.8 $\pm$ 3.5 & 95.8 $\pm$ 4.0 \\
$a\,\sin\,i$ (R$_\odot$) & 34.8 $\pm$ 1.5 & 40.3 $\pm$ 1.7 \\
$q\,=\,m_\mathrm{1}/m_\mathrm{2}$ & \multicolumn{2}{c}{1.16 $\pm$ 0.06} \\
$m\,\sin^3i$ (M$_\odot$)  & 6.4 $\pm$ 0.6 & 5.5 $\pm$ 0.5 \\
R$_\mathrm{RL}$\,$\sin\,i$ (R$_\odot$) & 13.6 $\pm$ 0.2 & 14.8 $\pm$ 0.2 \\
\hline
\end{tabular}
\end{center}
\end{table}
The rather low values of the RV amplitude ($K$), along with the modest values of $m\,\sin^3i$ as compared to the typical masses of stars of that spectral type, suggest that the system is seen under a rather low inclination angle, which would explain why the lines remain strongly blended over large parts of the orbital cycle. Consequently, Cyg\,OB2\,\#8A is unlikely to display eclipses. The projected Roche lobe radii (R$_\mathrm{RL}$\,$\sin\,i$, computed for an actual orbital separation equal to the semi-major axis) do not suggest any mass transfer through Roche lobe overflow.

We note that the accuracy of this orbital solution is affected by several factors. First, we measured RVs only for the \ion{He}{i} $\lambda$ 4471 line, because of the strong blending of all other lines. Second, our data do not allow us to state that a simple two-Gaussian model is adequate at all orbital phases. Indeed, individual lines can present intrinsic variability independent of their motion due to binarity. In that case, individual line profiles should not be gaussian anymore. Moreover, our data do not allow us to reject completely the possibility of a third component that could account for some discrepancies between the data and our orbital solution. More data obtained with a larger telescope, and with a better spectral resolution are necessary to address these issues. 

\section{Conclusions and future work \label{concl}} 
We present for the first time spectroscopic data revealing that Cyg\,OB2\,\#8A is an O6 + O5.5 binary system with a period of 21.9 d, likely seen under a rather low inclination angle. The two components are probably of different luminosity classes. The system appears to be eccentric, with a mass ratio of the two components close to 1. The binarity of Cyg\,OB2\,\#8A implies that the fundamental parameters derived through application of a single-star model atmosphere code (Herrero et al.\,\cite{HPN}) will have to be revised.

The \ion{He}{ii} $\lambda$ 4686 line presents a strong line profile variability which will be investigated in a subsequent paper to search for the signature of a putative wind-wind interaction.\\
The binarity of this system is crucial in the framework of the study of non-thermal radio emitters. In fact, 4 out of 8 O-stars displaying non-thermal radio emission listed by Rauw (\cite{nto}) are now confirmed binaries, while 2 more are suspected binaries. This lends further support to the idea that multiplicity and hence colliding winds are fundamental ingredients for the acceleration of the relativistic electrons (Eichler \& Usov \cite{EU}) that generate the synchrotron radio emission (White \cite{Wh}). {\it ASCA} data reveal that the X-ray spectrum of Cyg\,OB2\,\#8A shows a hard emission component with a temperature of more than 15\,$\times$\,10$^{6}$ K (De Becker \cite{DeBMT}). Provided that the thermal nature of this hard component is confirmed by our forthcoming observations with {\it XMM-Newton}, this temperature could be compatible with a plasma heated by a wind-wind collision (Stevens et al.\,\cite{SBP}).

However, the short period of this system raises questions regarding the detectability of synchrotron radiation in the radio domain. The radio photospheres, even at 3.6 cm, are huge (more than 1000 and 700 R$_\odot$ respectively for the primary and the secondary) and the collision zone of the two winds is deeply embedded therein. A similar situation occurs however in the case of WR\,104 (P $\sim$ 1 yr), where synchrotron radiation originates well within the dense wind of the Wolf-Rayet component (Monnier et al.\,\cite{Mon}). Considering this result, with a mass loss rate lowered by a factor of about 5-10 as expected for O stars compared to WR stars, even with a shorter separation, we expect that a non-thermal radio spectrum could be detected in the case of Cyg\,OB2\,\#8A. Information on the filling factors of the two winds (if they are clumpy) and on their geometry (if they are not spherical) are needed to understand this unexpected transparency. Additional studies are foreseen to address this issue, including an intense radio monitoring of Cyg\,OB2\,\#8A to search for a putative variability correlated with our orbital period.

Finally, as a binary system harbouring a small population of relativistic electrons, Cyg\,OB2\,\#8A may also contribute to the yet unidentified EGRET gamma-ray source 3EG J2033+4118 (Benaglia et al.\,\cite{Ben}) through inverse Compton scattering of photospheric UV photons. This hypothesis will be investigated through forthcoming observations of the Cyg\,OB2 region with the {\it INTEGRAL} satellite.  

\acknowledgement{We thank the referee Dr. Artemio Herrero for his careful reading and useful comments. We are greatly indebted to the FNRS (Belgium) for the financial support for the rent of the OHP telescope in 2000 and 2002 through contract 1.5.051.00 "Cr\'edit aux chercheurs". The travels to OHP were supported by the Minist\`ere de l'Enseignement Sup\'erieur et de la Recherche de la Communaut\'e Fran\c{c}aise. This research is also supported in part by contract PAI P5/36 (Belgian Federal Science Policy Office) and through the PRODEX XMM-OM Project. Our thanks go to Hugues Sana for helpful discussions on the orbital solution method.}

\end{document}